\documentclass[aps,prd,twocolumn,superscriptaddress,showpacs]{revtex4}

\usepackage{amsmath,amssymb}
\usepackage{graphicx}

\begin{document}


\title{
A numerical solution to the local cohomology problem in 
U(1) chiral gauge theories
}


\author{Daisuke Kadoh}
\email[]{kadoh@eken.phys.nagoya-u.ac.jp}
\author{Yoshio Kikukawa}
\email[]{kikukawa@eken.phys.nagoya-u.ac.jp}
\affiliation{Department of Physics, Nagoya University \ 
Nagoya 464-8602, Japan}

                                                                                        
\date{\today}

\begin{abstract}
We consider a numerical method to solve the local cohomology problem
related to the gauge anomaly cancellation
in U(1) chiral gauge theories.
In the cohomological analysis of the chiral anomaly, 
it is required to carry out the differentiation and the integration of the 
anomaly with respect to the continuous parameter for the interpolation of 
the admissible gauge fields. 
In our numerical approach,  the differentiation is evaluated explicitly 
through the rational approximation of the overlap Dirac operator 
with Zolotarev optimization. 
The integration is performed with a Gaussian Quadrature formula, which 
turns out to show rather good convergence. 
The Poincar\'e lemma is reformulated  for the finite lattice and 
is implemented numerically. We compute the current associated with the
cohomologically trivial part
of the chiral anomaly in two-dimensions and check its locality
properties.  

\end{abstract}


\maketitle


\section{\label{sec:intro} Introduction}

The construction of gauge-covariant and local lattice Dirac operators 
which satisfy the Ginsparg-Wilson relation\cite{Ginsparg:1981bj,
Neuberger:1997fp,Hasenfratz:1998ri,Neuberger:1998wv,
Hasenfratz:1998jp,Hernandez:1998et,Luscher:1998pq},
\begin{equation}
\gamma_5 D + D \gamma_5  = 2 a D \gamma_5 D ,
\end{equation}
has made it possible to introduce Weyl fermions on the lattice  and 
construct anomaly-free chiral gauge theories with exact gauge 
invariance\cite{Luscher:1998kn,Luscher:1998du,Luscher:1999un,
Luscher:1999mt,Luscher:2000hn}\cite{footnote:overlap}.
One of the crucial steps in the gauge-invariant construction
is to establish the exact cancellation of the gauge anomaly at a finite
lattice spacing.

In the case of U(1) chiral gauge 
theories\cite{Luscher:1998du}\cite{footnote:noncompact-u1},
the exact cancellation has been achieved through the cohomological 
classification of the chiral 
anomaly\cite{Luscher:1998kn,Fujiwara:1999fi,Fujiwara:1999fj}\cite{footnote:nonabelian-anomaly}. 
The anomaly is given in terms of 
lattice Dirac operator\cite{Luscher:1998pq,Kikukawa:1998pd, 
Adams:1998eg, Fujikawa:1998if, Suzuki:1998yz, Chiu:1998xf} as 
\begin{equation}
\label{eq:chiral-anomaly}
q(x) = \text{tr}\left\{ \gamma_5(1-a D)(x,x)\right\}
\end{equation}
and the local field 
$q(x)$ is a topological field 
in the sense that it satisfies 
\begin{equation}
\sum_x \delta q(x) = 0
\end{equation}
under a local variation of the gauge field.  
It follows from this property that the anomaly is 
cohomologically trivial, 
\begin{equation}
\label{eq:cohomological-triviality-of-chiral-anomaly}
\sum_\alpha e_\alpha q^\alpha(x) = \partial_\mu^\ast k_\mu(x), 
\quad\quad
q^\alpha(x) = \left. q(x) \right\vert_{U\rightarrow U^{e_\alpha}} ,
\end{equation}
for an anomaly-free multiplet of Weyl fermions which satisfies
the condition of the U(1) charges,
\begin{equation}
\sum_\alpha e_\alpha^3 = 0 . 
\end{equation}
Here $k_\mu(x)$ is a gauge-invariant local current. 
This  local current is in turn used in the gauge-invariant  construction of the 
functional measure of the Weyl fermions. 

If one thinks of the practical computation of observables in the lattice U(1) 
chiral gauge theories, it is required to compute 
the local current in Eq.~(\ref{eq:cohomological-triviality-of-chiral-anomaly})
for every admissible gauge field.  
The purpose of this paper is to attempt  a 
numerical computation of the local current. 
We will compute $k_\mu(x)$ in two-dimensions numerically
and  check its locality properties. 
We will also check how the exact cancellation of gauge anomaly 
works on the finite lattice. 

This paper is organized as follows. 
In section~\ref{sec:admissible-abelian-gauge-fields} 
we formulate the vector-potential-representation of the link variables  
for the admissible  U(1) gauge fields on the finite lattice. 
Using this representation, we introduce one-parameter families
of the admissible fields for the interpolations. 
In section~\ref{sec:topological-field} 
we describe our numerical method to 
compute the bi-local current which 
is the first-differential of the chiral anomaly with respect to
the vector potential.
In section~\ref{sec:cohomological-analysis-on-finite-lattice}
the Poincar\'e lemma is reformulated for a finite lattice
so that we can carry out the cohomological analysis directly on the finite lattice. 
The result of the cohomological analysis in two-dimensions is summarized. 
In section~\ref{sec:numerical-result-in-2D} we describe our numerical 
result of the computation of the local current $k_\mu(x)$.
Section~\ref{sec:discussion} is devoted to a summary and discussions. 

\section{\label{sec:admissible-abelian-gauge-fields}
Admissible U(1) gauge fields \ \ \ \ \ \ \ \ \ on a finite lattice}

Our first step is to formulate the vector-potential-representation of 
the link variables associated with an admissible U(1) gauge field
on a finite lattice.  Such a representation has been formulated
in the original cohomological analysis in \cite{Luscher:1998kn}. 
However, 
it is constructed for the admissible gauge fields on the infinite lattice and 
the resulted vector-potentials are not bounded in general. This representation 
therefore does not seem to be useful for numerical implementations. 
But, as has been shown in our previous paper\cite{Kadoh:2003ii}, 
it is possible to formulate the bounded and periodic vector-potential representation for the admissible gauge fields on the finite lattice.

We set the lattice spacing $a$ to unity and consider U(1) gauge fields 
on a finite two- or four-dimensional lattice ($n=2,4$) of size $L$ with periodic
boundary conditions.  $L$ is assumed to be an even integer for simplicity. 
The gauge fields on such a lattice can be represented through periodic
link fields on the infinite lattice, 
\begin{eqnarray}
&& U(x,\mu) \in {\rm U(1)}, \quad x=(x_1, x_2, \cdots, x_n) \in
\mathbb{Z}^n,
\\
&& U(x+L \hat \nu,\mu)=U(x,\mu) \ \  {\rm for \ all}  \ \
\mu,\nu=1,\cdots,n.
\end{eqnarray}
The independent degrees of freedom are then the link 
variables at the points in the region
\begin{equation}
\Gamma_n= \left\{x \in \mathbb{Z}^n \vert -L/2 \le x_\mu < L/2 \right\}.
\end{equation}
As to gauge transformations
\begin{equation}
U(x,\mu) \rightarrow \Lambda(x) U(x,\mu) \Lambda(x+\hat\mu)^{-1},
\end{equation}
we consider only periodic functions $\Lambda(x) \in $ U(1)
which preserves the periodicity of the link field. 

We impose the admissibility condition on the U(1)
gauge fields:
\begin{equation}
\vert F_{\mu\nu}(x) \vert  < \epsilon \ \  {\rm for \ all} \ \ 
x,\mu,\nu, 
\end{equation}
where the field tensor $F_{\mu\nu}(x)$ is defined through
\begin{eqnarray}
&& F_{\mu\nu}(x) = \frac{1}{i} {\rm ln} P_{\mu\nu}(x), 
\quad - \pi < F_{\mu\nu}(x) \le \pi ,
\\
&&
P_{\mu,\nu}(x)=U(x,\mu)U(x+\hat\mu,\nu)U(x+\hat\nu,\mu)^{-1}U(x,\nu)^{-1} . 
\nonumber\\
\end{eqnarray}
We require this condition because it ensures that 
the overlap Dirac 
operator\cite{Neuberger:1997fp,Neuberger:1998wv},
which we adopt in this work, 
is a smooth and local function of  the gauge field 
for $\vert 1-P_{\mu\nu}(x) \vert < 1/30$  in four-dimensions 
and $\vert 1-P_{\mu\nu}(x) \vert < 1/5$ in 
two-dimensions\cite{Hernandez:1998et}\cite{footnote:bound-neuberger}.
For $\epsilon < \pi/3$ 
the admissible U(1) gauge fields on the finite lattice
can be classified uniquely by the magnetic fluxes 
$m_{\mu\nu}$
(integers independent of $x$)
where
\begin{equation}
m_{\mu\nu} 
=
\frac{1}{2\pi}\sum_{s,t=0}^{L-1}F_{\mu\nu}(x+s\hat\mu+t\hat\nu).
\end{equation}
In this respect, the following field is periodic and can be shown
to have constant field tensor equal to $2\pi m_{\mu\nu}/L^2$:
\begin{equation}
\label{eq:link-of-magnetic-flux}
V_{[m]}(x,\mu) 
=
{\rm e}^{
-\frac{2\pi i}{L^2}\left[
L \delta_{\tilde x_\mu,L-1} \sum_{\nu > \mu} m_{\mu\nu}
\tilde x_\nu +\sum_{\nu < \mu} m_{\mu\nu} \tilde x_\nu\right]
},
\end{equation}
where the abbreviation $\tilde x_\mu = x_\mu$ mod $L$
has been used. Then any admissible U(1) gauge field
in the topological sector with the magnetic flux $m_{\mu\nu}$
may be expressed as 
\begin{equation}
\label{eq:separation-of-link}
U(x,\mu)=\tilde U(x,\mu) \, V_{[m]}(x,\mu) .
\end{equation}
We may regard $\tilde U(x,\mu)$ as 
the actual local and dynamical degrees of freedom 
in the given topological sector. 
This is because 
the magnetic flux $m_{\mu\nu}$ is invariant
with respect to a local variation of the link field.

The following lemma shows that 
it is possible to establish the one-to-one correspondence
between $\tilde U(x,\mu)$ and 
a periodic vector potential with the desired locality
properties on the finite lattice\cite{Kadoh:2003ii}. 

\vspace{1em}
\noindent{\bf Lemma \ref{sec:admissible-abelian-gauge-fields}} \ \ There
exists a periodic vector potential
$\tilde A_\mu(x)$  such that 
\begin{eqnarray}
&&   {\rm e}^{i \tilde A_\mu(x)}=\tilde U(x,\mu),\\
&&  \partial_\mu \tilde A_\nu(x) - \partial_\nu 
\tilde A_\mu(x) =F_{\mu\nu}(x) -\frac{2\pi m_{\mu\nu}}{L^2}, \\
&& 
\left\{ 
\begin{array}{ll}
\vert \tilde A_\mu(x) \vert \le \pi(1+4 \parallel x \parallel) &
 \parallel x \parallel \le L/2 \\
\vert \tilde A_\mu(x) \vert \le \pi(1+2L+ 2(n-1)L^2) & otherwise.
\end{array}
\right. \nonumber\\
\end{eqnarray}
\noindent Moreover, if $\tilde A_\mu^\prime(x)$ is any other
field with these properties, we have
\begin{equation}
\tilde A_\mu^\prime(x) = \tilde A_\mu(x)+ \partial_\mu 
\omega(x), 
\end{equation}
where the gauge function $\omega(x)$ takes values that are 
integer multiples of $2\pi$. 
The explicit formula of $\tilde A_\mu(x)$ in  
two-dimensions is given in the appendix. 

As emphasized in \cite{Luscher:1998kn}, 
it is important to note that
the locality properties of gauge invariant fields should be
the same independently of whether they are considered
to be functions of the link variables or the vector 
potential.  Since the mapping 
\begin{equation}
\tilde A_\mu(x)  \rightarrow \tilde U(x,\mu) = \text{e}^{i \tilde A_\mu(x)}
\end{equation}
is manifestly local, this is immediately clear if one starts with
a field composed from the link variables. In the other direction,
one may start from a gauge invariant local field $\phi(y)$ depending
the vector potential.  Then the key observation is that one is free
to impose a complete axial gauge taking the point $y$ as the origin. 
Around $y$ the vector potential is locally constructed from
the given link field and $\phi(y)$ thus maps to a local 
function of the link variables residing there. 

The vector potential $\tilde A_\mu(x)$
represents an admissible field through 
${\rm e}^{i \tilde A_\mu(x)} \times V_{[m]}(x,\mu)$ and 
the associated field tensor $F_{\mu\nu}(x) 
= \partial_\mu \tilde A_\nu(x)- \partial_\nu \tilde A_\mu(x)
+ \frac{2\pi m_{\mu\nu}}{L^2}$
is hence bounded by $\epsilon$. It is straightforward to 
check that this property is preserved if the potential is scaled
by a factor $t$ in the range $0 \le t \le 1$, i.e. we can  contract
the vector potential to zero without leaving the space of 
admissible fields. 
Then, in the cohomological analysis of the chiral anomaly, 
we may choose 
$V_{[m]}(x,\mu)$
as the reference gauge field in a given magnetic flux sector
and may consider the interpolation 
between the arbitrary gauge field in the same sector and
the reference field as follows:
\begin{equation}
\label{eq:u1-interpolation}
U_t(x,\mu)= {\rm e}^{i t \tilde A_\mu(x)} \times  V_{[m]}(x,\mu) 
\qquad t \in [0,1]. 
\end{equation}

\section{\label{sec:topological-field}
Chiral anomaly and its topological properties}

For our numerical application, 
we adopt the overlap Dirac operator\cite{Neuberger:1997fp,Neuberger:1998wv}
given by
\begin{equation}
D = \frac{1}{2}
\left(
1 + \gamma_5
\frac{H_{\rm w}}{\sqrt{H_{\rm w}^2}}
\right) , 
\end{equation} 
where $H_{\rm w}$ is the Hermitian Wilson-Dirac operator,
\begin{equation}
H_{\rm w}=\gamma_5 \left(
 \gamma_\mu \frac{1}{2}
( \nabla_\mu - \nabla_\mu^\dagger )
+ \frac{1}{2}\nabla_\mu \nabla_\mu^\dagger - m_0 \right)
\end{equation}
$ (0<m_0 <2)$.
It is a smooth and local function of  the admissible gauge field for 
$\vert 1-P_{\mu\nu}(x) \vert < 1/30$ in four-dimensions and
for $\vert 1-P_{\mu\nu}(x) \vert < 1/5$ in 
two-dimensions\cite{Hernandez:1998et}\cite{footnote:bound-neuberger}.
Then the chiral anomaly, Eq.~(\ref{eq:chiral-anomaly}), 
is given in terms of $H_{\rm w}$ as 
\begin{equation}
\label{eq:topological-field-in-overlapD}
q(x) 
= - \frac{1}{2}\text{tr}
\left\{ \frac{H_{\rm w}}{\sqrt{H_{\rm w}^2}}(x,x)\right\}. 
\end{equation}
As is clear from this expression, 
$q(x)$ is a topological field
with respect to any local variation 
of the admissible gauge field, 
\begin{equation}
 \sum_x  \delta q(x) = 0.
 \end{equation}

This topological property of the chiral anomaly can be cast into
properties of a certain gauge-invariant bi-local current. 
This bi-local current is defined through
the differentiation and integration with
respect to the continuous parameter $t$ for the interpolation as follows:
\begin{equation}
\label{eq:def-j}
j_\nu(x,y) = \int_0^1 dt \left(
\frac{\partial q(x)}{\partial 
\tilde A_\nu(y)} \right)_{\tilde A \rightarrow t \tilde A} .
\end{equation}
The original topological field can be expressed with the bi-local current as
\begin{equation}
\label{eq:chiral-anomaly-in-current}
q(x) = q_{[m]}(x) + \sum_{y \in \Gamma_n} 
j_\nu(x,y) \tilde A_\nu(y), 
\end{equation}
where $q_{[m]}(x)$ denotes the topological field for 
$V_{[m]}(x,\mu)$.   
Then the topological property and the gauge-invariance of 
$q(x)$ imply  that $j_\nu(x,y)$ satisfies 
\begin{equation}
\label{eq:topological-conditions-j}
\sum_{x \in \Gamma_n} j_\nu(x,y)=0, \qquad
 j_\nu(x,y)\overleftarrow{\partial_\nu^\ast} =0.
 \end{equation}
 These conditions provide the initial conditions for the 
 cohomological analysis. 
 
For our purpose, it is required to compute the above bi-local 
current numerically keeping the conditions Eq.~(\ref{eq:topological-conditions-j})
within a given accuracy. Our strategy is the following.
First of all, in order to obtain more explicit formula of  the bi-local current, 
we introduce the parameter representation of the inverse 
square root of $H_{\rm w}^2$ as 
\begin{eqnarray}
\frac{H_{\rm w}}{\sqrt{H_{\rm w}^2}}
&=& \int_{-\infty}^\infty \frac{dt}{\pi} \frac{H_{\rm w}}{t^2+ H_{\rm w}^2} . 
\end{eqnarray}
Then we can perform the differentiation of $q(x)$
with respect to the vector potential $\tilde A_\mu(x)$ 
explicitly\cite{Kikukawa:1998py} as
\begin{eqnarray}
\label{eq:del-q-del-A}
\frac{\partial q(x)}{\partial \tilde A_\nu(y)} 
&=& -\frac{1}{2}{\rm tr} \left\{
 \int_{-\infty}^\infty \frac{dt}{\pi}
 \frac{1} {t^2+H_{\rm w}^2} \times
 \right.
 \nonumber\\
 && 
  \left. 
 \left( t^2V_\mu(y)-H_{\rm w} V_\mu(y) H_{\rm w}
 \right) 
\frac{1} {t^2+ H_{\rm w}^2}
(x,x)
\right\} ,  \nonumber\\
\end{eqnarray}
where
\begin{eqnarray}
V_\mu(y)&=&-i\gamma_5 \frac{1}{2}
\left\{ 
(1-\gamma_\mu) \delta_{x,y} \delta_{x+\hat \mu, z}U_\mu(x) 
 \right.
 \nonumber\\
 && \qquad 
 \left.
\, +(1+\gamma_\mu) \delta_{y, z}\delta_{x,z+\hat \mu}U_\mu(z) ^{-1}
\right\} . \nonumber\\
\end{eqnarray}
For the numerical evaluation of the differentiation, 
we then adopt  the rational approximation
of the overlap Dirac operator\cite{Neuberger:1998my,Edwards:1998yw} 
with the Zolotarev 
optimization\cite{Chiu:2002eh,vandenEshof:2002ms}.
Namely, $q(x)$ is approximated by the formula with the degree $N_r$:
\begin{equation}
q(x) \fallingdotseq -\frac{1}{2} {\rm tr}\left\{ 
h_{\rm w} 
 \sum_{k=1}^{N_r} \frac{b_k}{h_{\rm w}^2 + c_{2k-1}} (x,x) 
\right\} . 
 \end{equation}
$h_{\rm w}$ is defined by $H_{\rm w}/ \lambda_{\rm min}$ where 
$\lambda_{\rm min}$ is the square root of the minimum of the
eigenvalues of $H_{\rm w}^2$. 
The definitions of the coefficients $c_k$ and $b_k$  are given in the appendix. 
With this approximation, Eq.~(\ref{eq:del-q-del-A}) 
can be approximated as follows:
\begin{eqnarray}
\label{eq:del-q-del-A-rational}
\frac{\partial q(x)}{\partial \tilde A_\nu(y)} 
&\fallingdotseq& -\frac{1}{2}{\rm tr} \left\{
 \sum_{k=1}^{N_r} b_k
 \frac{1} {h_{\rm w}^2 + c_{2k-1} } \times
 \right.
 \nonumber\\
 && 
  \left. 
 \left( c_{2k-1} v_\mu(y)-h_{\rm w} v_\mu(y) h_{\rm w}
 \right) 
\frac{1} {h_{\rm w}^2 + c_{2k-1}}
(x,x)
\right\} ,  \nonumber\\
\end{eqnarray}
where $v_\mu(y)=V_\mu(y) / \lambda_{\rm min}$.   
Finally,  for the integration back with respect to the parameter 
for the interpolation in Eq.~(\ref{eq:def-j}),  we adopt the 
Gaussian Quadrature (Gauss-Legendre) formula with  $N_g$ points:
\begin{equation}
\label{eq:approximation-j}
j_\nu(x,y) \fallingdotseq \sum_{i=1}^{N_g} w_i \left(
\frac{\partial q(x)}{\partial 
\tilde A_\nu(y)} \right)_{\tilde A \rightarrow t_i \tilde A} , 
\end{equation}
where $\{(t_i, w_i) | i=1,\cdots,N_g\}$ is the set of 
the abscissas and weights of the degree $N_g$. 

The formulae, Eqs.~(\ref{eq:del-q-del-A-rational}) and (\ref{eq:approximation-j}), 
are  still complicated for the numerical computation. 
But, in two-dimensions, it is not demanding numerically 
for relatively small lattice sizes 
to diagonalize $H_{\rm w}$ and store all the eigenvectors.
Namely,  we can write
\begin{equation}
h_{\rm w} (x,y)= \sum_{\lambda} \psi_\lambda(x)
\left( \frac{\lambda}{\lambda_{\rm min}}  \right) \psi_\lambda^\dagger(y) 
\end{equation}
and then it is possible to evaluate 
Eqs.~(\ref{eq:del-q-del-A-rational}), (\ref{eq:approximation-j})
explicitly. 

In this approximation, the gauge-invariance of the 
bi-local current and the second property of Eq.~(\ref{eq:topological-conditions-j})
can be preserved exactly. 
As to the first property of Eq.~(\ref{eq:topological-conditions-j}), 
we will see below that
the choice $N_r=18$ and $N_g=20$ gives 
good convergences for the admissible gauge fields on the two-dimensional 
lattice of the size $L=8,10,12$ and the original topological field, $q(x)$, can be 
reproduced through  Eq.~(\ref{eq:chiral-anomaly-in-current}) 
within the error less than $1 \times 10^{-14}$.

\section{\label{sec:cohomological-analysis-on-finite-lattice}
 Cohomological Analysis of chiral anomaly on a finite lattice}

\subsection{\label{subsec:lemma-on-finite-volume}
The modified Poincar\'e lemma on a finite lattice}
 
Our numerical cohomological analysis of the  chiral anomaly will be
performed directly on a finite lattice. 
For this purpose
it is required to reformulate the Poincar\'e lemma on the
lattice\cite{Luscher:1998kn} 
for a finite lattice. 
For the detail of 
the proof of the lemmas, refer to our previous 
paper\cite{Kadoh:2003ii}. 

In the following we will consider tensor fields 
$f_{\mu_1\cdots\mu_k}(x)$ on $\Gamma_n$ that are totally 
anti-symmetric in the indices $\mu_1,\cdots,\mu_k$. Such 
tensor fields 
may be regarded as periodic tensor fields on the infinite 
lattice,
\begin{equation}
f_{\mu_1\cdots\mu_k}(x+L\hat\nu)=f_{\mu_1\cdots\mu_k}(x) \ \ 
{\rm for \ all} \ \ \mu,\nu=1,\cdots, n.
\end{equation}
The locality properties of such fields are assumed to be as 
follows: for a certain reference point $x_0 \in \Gamma_n$ 
and $-L/2 \le (x_\mu-{x_0}_\mu) < L/2$ (mod $L$),
\begin{eqnarray}
\vert f_{\mu_1\cdots\mu_k}(x) \vert
&<& C_1 (1+\parallel x-x_0 \parallel^{p_1} ) \,  
{\rm e}^{-\parallel x-x_0\parallel/\varrho}, \nonumber\\
&& \qquad \qquad \quad 
\left( \parallel x-x_0 \parallel < L/2 \right), \\
\vert f_{\mu_1\cdots\mu_k}(x) \vert
&<& C_2 L^{p_2} \,  {\rm e}^{-L/2\varrho}, \nonumber\\
&&
\qquad \qquad \quad 
\left(\parallel x-x_0 \parallel \ge L/2 \right).
\end{eqnarray}
$\varrho$ is a localization range of the tensor field.
$C_i$ and $p_i \ge 0$ are certain constants that do not depend on $L$.
$\parallel x-x_0 \parallel $
is the taxi driver distance from $x_0$  to $x$.
This locality properties hold true for the differentials of 
the chiral anomaly given in terms of overlap Dirac 
operator\cite{Neuberger:1997fp,Neuberger:1998wv} 
with respect
to the admissible gauge 
fields\cite{Hernandez:1998et}\cite{footnote:lemma-for-superlocal-field}.

The differential forms on the finite lattice are 
introduced as in the continuum, following \cite{Luscher:1998kn}. 
If we adopt the Einstein summation convention for tensor indices,
the general $k$-form on $\Gamma_n$ is then given by 
\begin{equation}
f(x) = \frac{1}{k!} f_{\mu_1\cdots\mu_k}(x) 
dx_{\mu_1}\cdots dx_{\mu_k}.
\end{equation}
The linear space of all these forms is denoted
by $\Omega_k$.
An exterior difference operator ${\rm d} : \Omega_k 
\rightarrow \Omega_{k+1}$ may now be defined through
\begin{equation}
\label{eq:def-k-form}
d f(x) = \frac{1}{k!} \partial_\mu 
f_{\mu_1,\cdots,\mu_k}(x) dx_\mu dx_{\mu_1}\cdots dx_{\mu_k},
\end{equation}
where $\partial_\mu$ denotes the forward nearest-neighbor
difference operator. 
The associated divergence operator $d^\ast : \Omega_k 
\rightarrow \Omega_{k-1}$ is defined in the obvious way
by setting $d^\ast f =0 $ if $f$ is a 0-form and
\begin{equation}
d^\ast f(x) = \frac{1}{(k-1)!} \partial_\mu^\ast
f_{\mu\mu_2\cdots\mu_k}(x) dx_{\mu_2}\cdots dx_{\mu_k}
\end{equation}
in all other cases, where $\partial_\mu^\ast$ is the backward
nearest-neighbor difference operator.

By definition, the divergence operator satisfies that 
${d^\ast}^2=0$ and therefore 
the difference equation $d^\ast f = 0$ is solved by all
forms $f = d^\ast g$.
It has been shown that in the infinite lattice
these are in fact all solutions, 
an exception being the $0$-forms where one has a one-dimensional 
space of further solutions\cite{Luscher:1998kn}. 
This result is the lattice counter part of the Poincar\'e lemma known 
in the continuum theory. 
On the finite periodic lattice 
the lemma does not hold true any more, 
because the lattice is a n-dimensional torus and 
its cohomology group is now non-trivial. 
However,  
the lemma can be reformulated
so that it holds true up to exponentially small correction terms 
of order ${\cal O}({\rm e}^{-L/2\varrho})$ and 
for the form satisfying $\sum_{x \in \Gamma_n} f(x)=0$, 
it holds exactly
even on the finite lattice.\cite{footnote:lemma-for-superlocal-field}
The latter result is the lattice counter part of the 
corollary of de Rham theorem known in the continuum theory. 
The precise statements are the following.\cite{footnote:lemma-in-d}

\vspace{1em}
\noindent{\bf Lemma \ref{sec:cohomological-analysis-on-finite-lattice}.a} 
\ \ ({\bf Modified Poincar\'e lemma})

Let
$f$ be a $k$-form which satisfies
\begin{equation}
\label{eq:lemma-ast-condition}
d^\ast f = 0 \ \ \ {\rm and} \ \ \  \sum_{x\in \Gamma_n} f(x) =0 \ \ {\rm
if } \ \ k=0. 
\end{equation}
Then there exist a form $g \in \Omega_{k+1}$ and 
a form $\Delta f \in \Omega_{k}$ such that
\begin{equation}
f = d^\ast g + \Delta f , \qquad
\vert \Delta f_{\mu_1\cdots\mu_k}(x) \vert 
< c L^\sigma {\rm e}^{-L/2\varrho}.
\end{equation}

\vspace{1em}
\noindent{\bf Lemma \ref{sec:cohomological-analysis-on-finite-lattice}.b}
\ \ ({\bf Corollary of de Rham theorem})

Let
$f$ be a $k$-form which satisfies
\begin{equation}
\label{eq:lemma-ast-condition-b}
d^\ast f = 0 \ \  {\rm and} \ \  \sum_{x\in \Gamma_n} f(x) =0 .
\end{equation}
Then there exist a form $g \in \Omega_{k+1}$ 
such that
\begin{equation}
f = d^\ast g .
\end{equation}

\vspace{1em}
\noindent
The explicit formula of $g(x)$ in two-dimensions ($n=2$) is given
in the appendix.

\subsection{\label{subsec:solution-in-2d}
A solution to the local cohomology problem \\
on the two-dimensional finite lattice}

In two dimensions, the cohomological analysis results in 
the following formula for $q(x)$:
\begin{eqnarray}
\label{eq:chiral-anomaly-in-phi-prime}
&&q(x)=
{\cal A}(x)
+ \partial_\mu^\ast h_\mu(x)
+ \Delta q(x),   
\end{eqnarray}
where
\begin{equation}
{\cal A}(x)=q_{[m]}(x)+
\phi_{\mu\nu}(x) \,  \tilde F_{\mu\nu}(x), \quad
\partial_\mu^\ast \phi_{\mu\nu}(x) = 0 . 
\end{equation}
$\phi_{\mu\nu}(x)$
is obtained through the applications of
the modified Poincar\'e lemma and the corollary of 
the de Rham theorem:
\begin{eqnarray}
\label{eq:phi}
j_\nu(x,y) &\xrightarrow{\text{IV.a}}&
\theta_{\nu\mu}(x,y) \overleftarrow{\partial_\mu^\ast} 
+ \Delta j_\nu(x,y), 
\\
\sum_{z \in \Gamma_2}  \Delta j_\nu(z,x)
&\xrightarrow{\text{IV.b}}& -2 \, \Delta \Phi_{\nu \mu}(x)
  \overleftarrow{\partial_\mu^\ast} 
\end{eqnarray}
and 
\begin{eqnarray}
\phi_{\mu\nu}(x) 
&\equiv&
\frac{1}{2}\sum_{z \in \Gamma_2}
\theta_{\mu\nu}(z,x) 
-\Delta \Phi_{\mu\nu}(x).  
\end{eqnarray}
In two-dimensions, $\phi_{\mu\nu}(x) $ is a 
constant and assumes the form $\phi_{\mu\nu}(x) = \gamma_{[m,w]} 
\epsilon_{\mu\nu}$. 
$h_\mu(x)$ is obtained through the application of 
the modified Poincar\'e lemma:
\begin{eqnarray}
\theta_{\mu\nu}(x,y) - 
   \delta_{x,y} \sum_{z \in \Gamma_2} \theta_{\mu\nu}(z,y) 
&\xrightarrow{\text{IV.a}}&\partial_\lambda^\ast \tau_{\lambda\mu\nu}(x,y)
\nonumber\\
\end{eqnarray}
and
\begin{eqnarray}
h_\mu(x) &\equiv& \frac{1}{2}\sum_{y \in \Gamma_2} 
\tau_{\mu\nu\rho}(x,y) \tilde F_{\nu\rho}(y). 
\end{eqnarray}
$\Delta q(x)$ is defined by 
\begin{eqnarray}
\Delta q(x) &\equiv&\Delta \Phi_{\mu\nu}(x) \tilde F_{\mu\nu}(x)
+\sum_{y \in \Gamma_2} 
\Delta j_\nu(x,y) \tilde A_\nu(y).  \nonumber\\
\end{eqnarray}

For the anomaly-free multiplet 
satisfying the condition $\sum_\alpha e_\alpha^2 =0$, 
the anomaly part ${\cal A}(x)=q_{[m]}(x)+\gamma_{[m]} 
\epsilon_{\mu\nu} \tilde F_{\mu\nu}(x)$ cancels 
up to an exponentially small term,
\begin{equation}
\sum_\alpha e_\alpha {\cal A}^\alpha(x) = 
\Delta {\cal A}(x) ,
\end{equation}
where ${\cal A}^\alpha(x) = \left. {\cal A}(x) \right\vert_{U\rightarrow U^{e_\alpha}} $.
Since $\sum_{x \in \Gamma_2} \Delta {\cal A}(x)  =0$, we may 
apply the modified Poincar\'e lemma IV.a to obtain 
\begin{eqnarray}
\label{eq:anomaly-part-current}
\Delta {\cal A}(x) 
&\xrightarrow{\text{IV.a}} &
\partial_\mu^\ast  \Delta k_\mu(x) .
\end{eqnarray}
On the other hand, since $\sum_{x \in \Gamma_2}  \Delta q(x) =0$, we may 
also apply the lemma IV.b as follows:
\begin{eqnarray}
\label{eq:finite-volume-correnctiona-current}
\Delta q(x) 
&\xrightarrow{\text{IV.b}} &
 \partial_\mu^\ast  \Delta h_\mu(x) .
\end{eqnarray}
Then we can see that $q(x)$ is cohomologically trivial,
\begin{equation}
\sum_\alpha e_\alpha q^\alpha(x) = \partial_\mu^\ast k_\mu(x),
\end{equation}
where $k_\mu(x)$ is given explicitly as 
\begin{equation}
\label{eq:final-formula-kmu}
k_\mu(x)\equiv \sum_\alpha e_\alpha 
\{ h_\mu(x)+\Delta h_\mu(x) \}^\alpha
+\Delta k_\mu(x). 
\end{equation}

Therefore, once the bi-local current $j_\mu(x,y)$ is computed, 
the current $k_\mu(x)$ can be obtained by the sequence of the 
applications of the lemmas IV.a,b. 
The numerical implementation of this step is straightforward
using the explicit solutions,  $g(x)$, of the lemmas
given in the appendix. 

\section{\label{sec:numerical-result-in-2D} Numerical results 
}

We now describe our result of numerical computations of the 
local current $k_\mu(x)$.  We consider the lattice sizes $L=8,10,12$. 
Admissible gauge fields are generated by Monte Carlo simulation 
using the action 
\begin{equation}
\label{eq:gauge-action}
S=\frac{1}{4e^2} \sum_{x,\mu\nu} 
\frac{F_{\mu\nu}(x)^2}{1-F_{\mu\nu}(x)^2/ \epsilon^2}
 = \beta \sum_{\square}\frac{F_{\square}^2}{1-F_{\square}^2/ \epsilon^2} . 
\end{equation}
As reported in \cite{Fukaya:2003ph}, the topological charge is preserved 
during the Monte Carlo updates with this type of action, even when 
$\epsilon$ is set to $\pi$.  We adopt this option
and check the locality of the topological field numerically
for several values of $\beta$.  
We consider the topological sectors with $m_{12}=0, 1$ and 
the initial configuration is chosen as $V_{[m]}(x,\mu)$ with a given $m_{12}$. 

For each admissible gauge field, we compute the vector potential $\tilde A_\mu(x)$ 
formulated in section~\ref{sec:admissible-abelian-gauge-fields}. 
We also compute the abscissas and weights $\{(t_i, w_i) \vert \ i=1,\cdots, N_g\}$ 
for Gaussian Quadrature formula with the degree $N_g$. 
By this, the discrete interpolation of the given admissible gauge field is fixed: 
\begin{equation}
\left\{ U^{(i)}(x,\mu)={\rm e}^{i  t_i \tilde A_\mu(x)} \times V_{[m]}(x,\mu) \
 \big\vert \ \ i=1, \cdots, N_g \right\}. 
\end{equation}
In table~\ref{tab:gauss-legendre}, the abscissas and weights 
for the degree $N_g=20$ are shown. 
\begin{table}[htdp]
\caption{\label{tab:gauss-legendre} 
The abscissas and weights for the
Gauss-Legendre formula are shown for $N_g=20$.  $t \in [0,1]$.  
}

\begin{ruledtabular}
\begin{tabular}{c|c|c}
   & $t_i$  & $w_i$ \\ \hline 
   1  & 3.435700407452558E-003  & 8.807003569576136E-003 \\
   2  & 1.801403636104310E-002  & 2.030071490019346E-002 \\
   3  & 4.388278587433703E-002  & 3.133602416705449E-002 \\
   4  & 8.044151408889061E-002  & 4.163837078835234E-002 \\
   5  & 0.126834046769925  & 5.096505990862024E-002 \\
   6  & 0.181973159636742  & 5.909726598075923E-002 \\
   7  & 0.244566499024586  & 6.584431922458825E-002 \\
   8  & 0.313146955642290  & 7.104805465919108E-002 \\
   9  & 0.386107074429177  & 7.458649323630189E-002 \\
   10  & 0.461736739433251  & 7.637669356536294E-002 \\
   11  & 0.538263260566749  & 7.637669356536294E-002 \\
   12  & 0.613892925570823  & 7.458649323630189E-002 \\
   13  & 0.686853044357710  & 7.104805465919108E-002 \\
   14  & 0.755433500975414  & 6.584431922458825E-002 \\
   15  & 0.818026840363258  & 5.909726598075923E-002 \\
   16  & 0.873165953230075  & 5.096505990862024E-002 \\
   17  & 0.919558485911109  & 4.163837078835234E-002 \\
   18  & 0.956117214125663  & 3.133602416705449E-002 \\
   19  & 0.981985963638957 &  2.030071490019346E-002 \\
   20  & 0.996564299592547  & 8.807003569576136E-003 \\
\end{tabular}

\end{ruledtabular}

\end{table}%

For each $U^{(i)}(x,\mu)$, all eigenvalues and eigenvectors of $H_{\rm w}$ 
are computed numerically using the Householder method.  
We choose $m_0=0.9$ in $H_{\rm w}$. 
Then we compute the bi-local current 
$j_\nu(x,y)$ through Eqs.~(\ref{eq:del-q-del-A-rational}) 
and (\ref{eq:approximation-j}). 

We can check the convergence of the above procedure
through Eq.~(\ref{eq:chiral-anomaly-in-current}) 
by computing the deviation
\begin{equation}
\chi = {\rm max}_{x \in \Gamma_2} 
\left\vert \delta(x) \right\vert, 
\end{equation}
where 
\begin{equation}
\delta (x) =q(x) - q_{[m]}(x) - \sum_{y \in \Gamma_2} 
j_\nu(x,y) \tilde A_\nu(y) .
\end{equation}
Here the original topological fields $q(x)$ and $q_{[m]}(x)$ are constructed
by computing all eigenvalues and eigenvectors of $H_{\rm w}$ 
for the given admissible gauge field $U(x,\mu)$ and for the reference gauge field
$V_{[m]}(x,\mu)$, respectively.  In this computation, we found that
the topological fields have typically the values of order $O(10^{-2})-O(10^{-3})$. 
The integer topological charge $Q=\sum_{x \in \Gamma_2} q(x)=m_{12}$ 
is reproduced within the error of order $O(10^{-12})-O(10^{-13})$.
In table~\ref{tab:j-convergence}, 
we show the dependence of $\chi$ on the degree $N_g$ 
with $N_r=18$, $L=8$ and $\beta=3.0$  fixed. 
\begin{table}[htdp]
\caption{\label{tab:j-convergence} 
The convergence of the bi-local current is shown for 
various values of $N_g$ with $N_r=18, L=8, \beta=3.0$ fixed.}

\begin{ruledtabular}
\begin{tabular}{ccr}
 $N_g$ & $ \chi$ \\
 \hline 
  4 &  $ O( 10^{-  8} ) $ & \\
  6 &  $ O( 10^{-12} ) $ & \\
  8 &  $ O( 10^{-14} ) $ & \\
  10 & $O( 10^{-15} ) $ & \\
  12 & $O( 10^{-15} ) $ &  \\
  16 & $O( 10^{-15} ) $ & \\
\end{tabular}

\end{ruledtabular}

\end{table}%

\noindent
For the larger lattice sizes $L=10,12$, we found that $\chi$ is  less than 
$1 \times 10^{-14}$ with the choice of $N_g=20$ and $N_r=18$. 

Once the bi-local current $j_\mu(x,y)$ is computed, 
the cohomological analysis of the chiral anomaly
can be performed  numerically by the sequence of the 
applications of the lemmas.  The result can be expressed as
(cf. Eq.~(\ref{eq:chiral-anomaly-in-phi-prime}) )
\begin{eqnarray}
\label{eq:chiral-anomaly-in-A-h-dq}
&&q(x)=
{\cal A}(x)
+ \partial_\mu^\ast h_\mu(x)
+ \Delta q(x),   
\end{eqnarray}
where
\begin{equation}
{\cal A}(x)=q_{[m]}(x)+
\gamma_{[m,w]} 
\epsilon_{\mu\nu}\,  \tilde F_{\mu\nu}(x) .
\end{equation}

In order to check the locality properties of the fields, $q(x)$,  
${\cal A}(x)$, $h_\mu(x)$ and $\Delta q(x)$, we apply 
a small local variation to the given admissible gauge field as
$ U(x,\mu) \rightarrow {\rm e}^{i \eta_\mu(x)} U(x,\mu) $ where 
\begin{equation}
\eta_\mu(x) = 0.05 \times 2\pi \, \delta_{x,x_0} \delta_{\mu,1} 
\end{equation}
and compute the variations of the fields. 
For each variation of the fields, $\delta_\eta f(x)$, we define
\begin{equation}
\delta_\eta f(r) =  \text{max} \left\{ \vert  \delta_\eta f(x) \vert \big\vert 
\, r=\parallel x-x_0 \parallel \right\}   
\end{equation}
and see the locality properties of the fields by 
plotting $\delta_\eta f(r)$  against $r=\parallel x-x_0 \parallel$. 
For the anomaly part ${\cal A}(x)$,  
the variation of $\gamma_{[m,w]}$ is considered. 

\begin{figure}[htbp]
\begin{center}
\includegraphics[width=9.5cm]{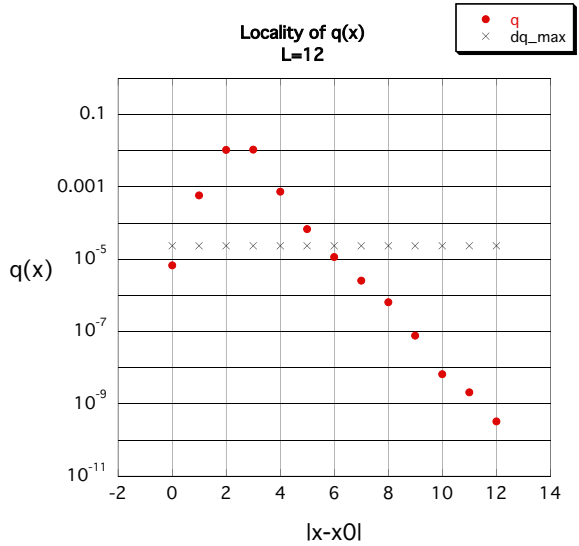}
\caption{\label{fig:locality12q} 
The variation of the topological field
$\delta_\eta q(r)$(filled circle) is plotted against 
$r=\parallel x-x_0 \parallel$. 
The maximum value of $\vert \Delta q(x) \vert$ (cross) is also shown.
The lattice size is $L=12$. The gauge field is generated at $\beta=3.0$
with the vanishing magnetic flux $m_{12}=0$.
 }
\end{center}
\end{figure}

\begin{figure}[htbp]
\begin{center}
\includegraphics[width=9.5cm]{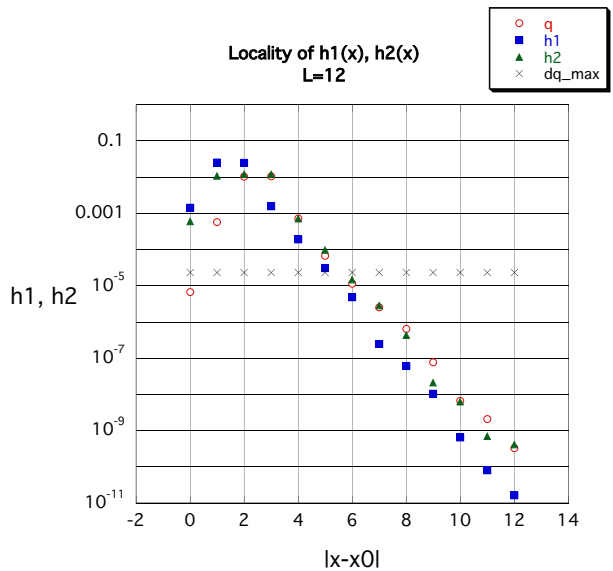}
\caption{\label{fig:locality12h} 
The variations 
$\delta_\eta h_\mu(r)$ are plotted against 
$r=\parallel x-x_0 \parallel$ (filled square and triangle). 
For a guide, the variation $\delta_\eta q(r)$ (open circle) and 
the maximum value of $\vert \Delta q(x) \vert$(cross) are also shown.
}
\end{center}
\end{figure}

\begin{figure}[htbp]
\begin{center}
\includegraphics[width=9.5cm]{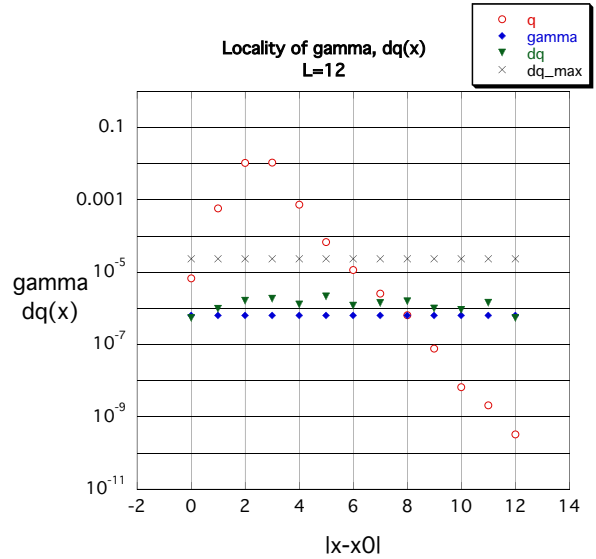}
\caption{\label{fig:locality12g} 
The variations 
$\delta_\eta \gamma_{[m,w]}$(filled diamond) and 
$\delta_\eta \Delta q(r)$ (filled triangle)
are plotted against 
$r=\parallel x-x_0 \parallel$ . 
For a guide, the variation $\delta_\eta q(r)$ (open circle) and 
the maximum value of $\vert \Delta q(x) \vert$(cross) are also shown.
}
\end{center}
\end{figure}

The results are shown in figures~\ref{fig:locality12q}, \ref{fig:locality12h} and 
\ref{fig:locality12g} for $\beta=3.0$. 
In figure~\ref{fig:locality12q}, the variation of the topological field $q(x)$ 
is shown. 
The locality of the topological field 
$q(x)$ is clearly seen.  We can read the locality range 
as $\varrho \simeq 0.5$. 
The maximum value of the field $\Delta q(x)$ is also shown in the same
figure. We can confirm that
\begin{equation}
\vert \Delta q(x) \vert  
\simeq O(10^{-5}) 
\simeq O ( {\rm e}^{-L/2\varrho}) 
\quad
(L=12). 
\end{equation} 
In figure~\ref{fig:locality12h}, the variations of the current $h_\mu(x)$
are shown. It shows clearly that the current $h_\mu(x)$ has the same
locality property as the topological field $q(x)$ has and thus is local. 
In figure~\ref{fig:locality12g}, the variations of the anomaly coefficient 
$\gamma_{[m,w]}$ and the field $\Delta q(x)$ are shown. 
We can confirm that 
the gauge field dependence of $\gamma_{[m,w]}$ is indeed small
and the same order of magnitude as the size of $\Delta q(x)$ and its variation. 

\begin{figure}[htbp]
\begin{center}
\includegraphics[width=9.5cm]{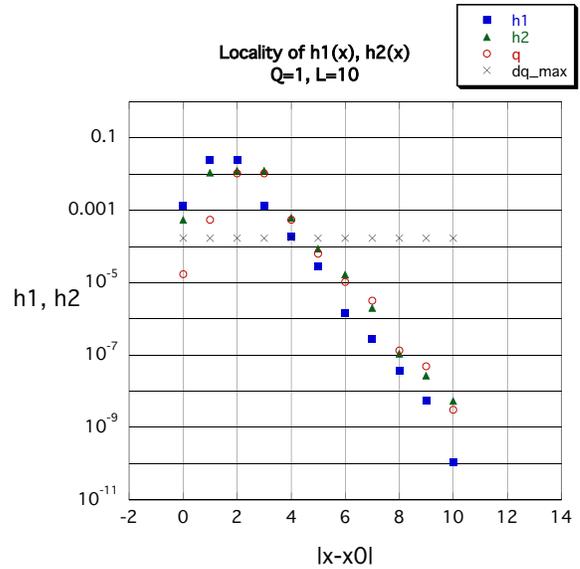}
\caption{\label{fig:locality10hm=1} 
The variations 
$\delta_\eta h_\mu(r)$ are plotted against 
$r=\parallel x-x_0 \parallel$ (filled square and triangle). 
For a guide, the variation $\delta_\eta q(r)$ (open circle) and 
the maximum value of $\vert \Delta q(x) \vert$(cross) are also shown.
The lattice size is $L=10$. The gauge field is generated at $\beta=3.0$
with the magnetic flux $m_{12}=1$.
}
\end{center}
\end{figure}

The locality properties of the fields $q(x)$,  
${\cal A}(x)$, $h_\mu(x)$ and $\Delta q(x)$ are 
confirmed also for topologically non-trivial gauge fields. 
See 
figure~\ref{fig:locality10hm=1}. 

\begin{figure}[htbp]
\begin{center}
\includegraphics[width=9.5cm]{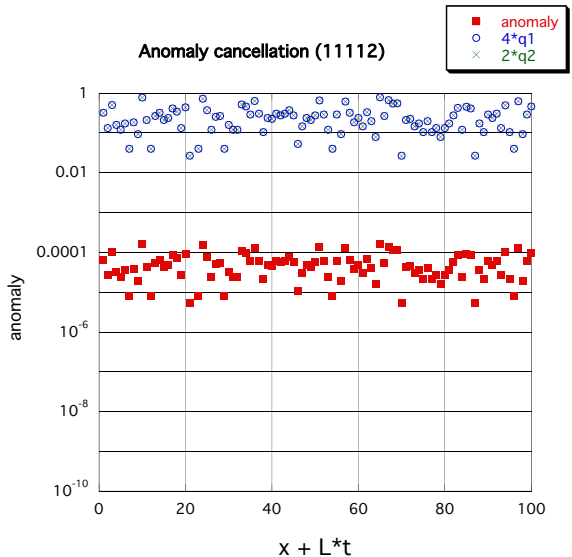}
\caption{\label{fig:anomaly10}
The anomaly parts of the Left-handed fermions $4\times {\cal  A}^1(x)$ (open circle)
and of the Right-handed fermion ${\cal  A}^2(x)$ (cross) are plotted against 
the fused coordinate $x+ L  \times t$.  The total anomaly part $4\times {\cal  A}^1(x)
- {\cal  A}^2(x)$ (filled rectangle) is also plotted. 
The lattice size is $L=10$. The gauge field is generated at $\beta=3.0$
with the vanishing magnetic flux $m_{12}=0$.
}
\end{center}
\end{figure}

We next examine the cancellation of the gauge anomaly. 
We consider the so-called 11112 model which consists
of four Left-handed Weyl fermions with unit charge
and one Right-handed Weyl fermion with charge two. 
The gauge anomaly cancellation condition in two-dimensions is satisfied as follows:
\begin{equation}
\sum_{i=1}^4  e^2 - (2e)^2 = 0.
\end{equation}
In figure~\ref{fig:anomaly10}, we plot the anomaly parts of the Left-handed 
fermions $4 \times {\cal A}^1(x)$ and of the Right-handed fermion 
$ {\cal A}^2(x)$,
where 
${\cal A}^\alpha(x) = \left. {\cal A}(x) \right\vert_{U\rightarrow U^{e_\alpha}} $,
and the total anomaly part 
$\sum_\alpha {\cal A}^\alpha(x)=4\times {\cal A}^1(x) - {\cal A}^2(x)$
for $\beta=3.0$ and $m_{12}=0$. 
The result is impressive.  The size of the total anomaly part is reduced to 
the order $O({\rm e}^{-L/2\varrho})$ after the cancellation and we can confirm that 
\begin{equation}
 \Delta {\cal A}(x) 
\simeq O({\rm e}^{-L/2\varrho})
\simeq O( 10^{-4} ) \quad (L=10). 
\end{equation}

\begin{figure}[htbp]
\begin{center}
\includegraphics[width=9.5cm]{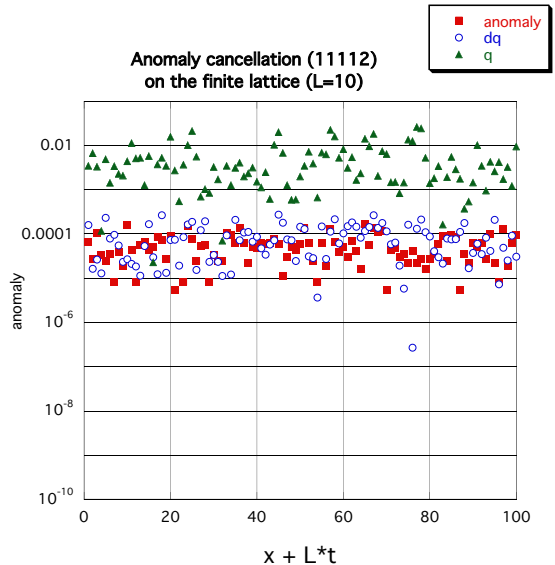}
\caption{\label{fig:cancellation10} 
The total topological field $\sum_\alpha q^\alpha(x)$ (filled triangle), 
the total anomaly part $\sum_\alpha {\cal  A}^\alpha(x)$ (filled rectangle)
and the total finite volume correction $\sum_\alpha \Delta q^\alpha(x)$ (open circle)
are plotted against the fused coordinate $x+ L  \times t$.  
}
\end{center}
\end{figure}

This result is also confirmed in figure~\ref{fig:cancellation10}. 
In this figure, the total topological field $\sum_\alpha q^\alpha(x)$, 
the total anomaly part $\sum_\alpha {\cal A}^\alpha(x)$ and 
the total finite volume correction $\sum_\alpha \Delta q^\alpha (x)$
are plotted.  We see clearly that 
the size of $\Delta {\cal A}(x)$ is the same order of magnitude
as the size of $\Delta q(x)$.  The similar cancellation is also observed 
for topologically non-trivial gauge fields with $m_{12}=1$. 

Now the current $k_\mu(x)$ can be computed from 
Eqs.~(\ref{eq:anomaly-part-current}), (\ref{eq:finite-volume-correnctiona-current})
and (\ref{eq:final-formula-kmu}). The current so obtained is thus local. 
This local current can reproduce the original topological charge $q(x)$ 
within the deviation of order $O(10^{-15})$. 
The gauge invariance of the current is also maintained within the error.

\section{\label{sec:discussion} Discussion}

We have demonstrated that the cohomological analysis
of the chiral anomaly associated with the overlap fermions
can be performed numerically in two-dimensional lattice with a finite volume. 
The resulted current $k_\mu(x)$ is gauge invariant and local. 

Four-dimensional case is numerically demanding.  
The evaluation of the differential of the topological fields
should be performed without diagonalizing $H_{\rm w}$. For other parts,
our procedure would  work also in this case. 

Our next step would be the numerical construction of the 
Weyl fermions measures and observables in the anomaly-free U(1)
chiral gauge theories in two-dimensions, keeping the gauge invariance. 
Work in this direction is in progress.

\appendix

\section{\label{app:vector-potential-in-axial-gauge}}
In this appendix, we give the explicit formulae for the 
vector-potential representation of the link variables
of the admissible U(1) gauge field discussed in 
the section \ref{sec:admissible-abelian-gauge-fields}. 
We first define 
\begin{equation}
\tilde a_\mu(x) \equiv \frac{1}{i} \ln \{ U(x,\mu) V_{[m]}(x,\mu)^{-1} \} 
\end{equation}
and 
\begin{equation}
2 \pi \tilde n_{\mu\nu}(x) \equiv 
F_{\mu\nu}(x) -\frac{2\pi m_{\mu\nu}}{L^2}
-\{ \partial_\mu \tilde a_\nu(x) - \partial_\nu \tilde a_\mu(x) \} . 
\end{equation}
We also introduce a summation convention as
\begin{equation}
\sum_{t_i =0}^{x_i-1}{}^\prime f(x)
=\left\{ 
\begin{array}{ll}
\sum_{t_i=0}^{x_i-1} f(x) & (x_i \ge 1 )\\
0 & (x_i =0 )\\
\sum_{t_i=x_i}^{-1} (-1) f(x) & (x_i \le -1 )
\end{array}
 \right. .
\end{equation}

Then, in four-dimensions, the vector-potential is defined by
\begin{eqnarray}
\tilde A_4(x) &=&  \tilde a_4(x) + 2\pi \left\{
\sum_{t_1=0}^{x_1-1}{}^\prime  \tilde n_{14}(x)
+\sum_{t_2=0}^{x_2-1}{}^\prime 2\pi \tilde n_{24}(x)\vert_{x_1=0}
\right.
\nonumber\\
&& 
\left. 
+\sum_{t_3=0}^{x_3-1}{}^\prime \tilde n_{34}(x)\vert_{x_1=x_2=0}
\right\}, 
\nonumber\\
\tilde A_3(x) &=&  \tilde a_3(x) + 2\pi  \left\{
\sum_{t_1=0}^{x_1-1}{}^\prime  \tilde n_{13}(x)
+\sum_{t_2=0}^{x_2-1}{}^\prime  \tilde n_{23}(x)\vert_{x_1=0}
\right.
\nonumber\\
&&
\left. 
-\delta_{x_3,L/2-1}\sum_{t_4=0}^{x_4-1}{}^\prime
\sum_{t_3=-L/2}^{L/2-1} \tilde n_{34}(x)\vert_{x_1=x_2=0} \right\}, 
\nonumber\\
\tilde A_2(x) &=&  \tilde a_2(x) + 2\pi \left\{ 
\sum_{t_1=0}^{x_1-1}{}^\prime \tilde n_{12}(x) 
\right. 
\nonumber\\
&&
-\delta_{x_2,L/2-1}\sum_{t_3=0}^{x_3-1}{}^\prime
\sum_{t_2=-L/2}^{L/2-1} \tilde n_{23}(x)\vert_{x_1=0,x_4=0} 
\nonumber\\ 
&&
\left.
 -\delta_{x_2,L/2-1}\sum_{t_4=0}^{x_4-1}{}^\prime
\sum_{t_2=-L/2}^{L/2-1} \tilde n_{24}(x)\vert_{x_1=0} \right\}, 
\nonumber\\
\tilde A_1(x) &=& \tilde a_1(x) + 2\pi \left\{ 
\phantom{
-\delta_{x_1,L/2-1} \sum_{t_2=0}^{x_2-1}{}^\prime
\sum_{t_1=-L/2}^{L/2-1} \tilde n_{12}(x)\vert_{x_3=x_4=0} 
}
\right. 
\nonumber\\
&&
-\delta_{x_1,L/2-1} \sum_{t_2=0}^{x_2-1}{}^\prime
\sum_{t_1=-L/2}^{L/2-1} \tilde n_{12}(x)\vert_{x_3=x_4=0} 
\nonumber\\ 
&& -\delta_{x_1,L/2-1} \sum_{t_3=0}^{x_3-1}{}^\prime 
\sum_{t_1=-L/2}^{L/2-1} \tilde n_{13}(x)\vert_{x_4=0} 
\nonumber\\
&&
\left.  -\delta_{x_1,L/2-1} \sum_{t_4=0}^{x_4-1}{}^\prime
\sum_{t_1=-L/2}^{L/2-1} \tilde n_{14}(x)
\right\}. \nonumber
\end{eqnarray}

In two-dimensions, the vector potential is defined by
\begin{eqnarray}
\tilde A_2(x) &=&  \tilde a_2(x) + 2\pi  
\sum_{t_1=0}^{x_1-1}{}^\prime \tilde n_{12}(x) ,
\nonumber\\
\tilde A_1(x) &=& \tilde a_1(x)   \nonumber\\
&&
-\delta_{x_1,L/2-1} \sum_{t_2=0}^{x_2-1}{}^\prime
\sum_{t_1=-L/2}^{L/2-1} 2\pi \tilde n_{12}(x) .
\nonumber\\ 
\end{eqnarray}

\section{\label{app:rational-approximation} Rational approximation of 
the Overlap Dirac operator}

In this appendix, we give the formula of 
the rational approximation of the overlap Dirac operator with the 
Zolotarev optimization. The inverse square root of $H_{\rm w}^2$ 
times $H_{\rm w}$ is approximated by 
 \begin{equation}
 \frac{H_{\rm w}}{\sqrt{H_{\rm w}^2}}
 = h_{\rm w} 
 \sum_{k=1}^n \frac{b_k}{h_{\rm w}^2 + c_{2k-1}} . 
 \end{equation}
 $h_{\rm w}$ is defined by $H_{\rm w}/ \lambda_{\rm min}$ where 
$\lambda_{\rm min}$ is the square root of the minimum of the
eigenvalues of $H_{\rm w}^2$, 
and the coefficients $c_k$ and $b_k$  are given as follows:
\begin{eqnarray}
c_k &=& \frac{{\rm sn}^2\left(\frac{k K^\prime}{2n}; \kappa^\prime \right)}
         {1-{\rm sn}^2\left(\frac{k K^\prime}{2n}; \kappa^\prime \right)}, \\
\lambda &=& \prod_{k=1}^{2n}
\frac{\Theta^2\left(\frac{2k K^\prime}{2n}; \kappa^\prime \right)}
       {\Theta^2\left(\frac{(2k-1) K^\prime}{2n}; \kappa^\prime \right)}, \\
d_0 &=& \frac{2 \lambda}{\lambda+1}\prod_{k=1}^n 
\frac{1+c_{2k-1}}{1+c_{2k}} , \\
b_k &=& d_0 \frac{\prod_{i=1}^{n-1}(c_{2i}-c_{2k-1})}
                             {\prod_{i=1, i \not = k}^{n}(c_{2i-1}-c_{2k-1})} .
\end{eqnarray}
$K^\prime$ is the complete elliptic integral of the first kind 
with modulus 
$\kappa^\prime=\sqrt{1-(\lambda_{\rm min}/\lambda_{\rm max})^2}$.

\section{\label{app:solution-of-lemma} Solutions of the Modified Poincar\'e lemma
and the corollary of de Rham theorem 
in two-dimensions}

In this appendix, we give the explicit solutions of the Poincar\'e lemma
in two-dimensions (n=2). 

IV.a(0-form) and IV.b(0-form):
\begin{eqnarray}
&& g_1(x_1,x_2) = \delta_{x_2,x^0_2} \sum_{y_1=x^0_1-L/2} ^{x_1} 
 \bar f(y_1), \nonumber\\
 && g_2(x_1,x_2)= \sum_{y_2=x^0_2-L/2} ^{x_2}
 \left\{ f(x_1,y_2)-\delta_{y_2,x^0_2} \bar f(x_1) \right\}, \nonumber\\
 \end{eqnarray}
 where
$\bar f(x_1) = \sum_{y_2=-L/2}^{L/2-1}  f(x_1,y_2)$.

\vspace{1em}
IV.a(1-form):
\begin{eqnarray}
&& g_{12}(x_1,x_2)=- \sum_{y_2=x^0_2-L/2}^{x_2}
\left\{ f_1(x_1,y_2) - \delta_{y_2,x^0_2} \bar f_1(x_1) \right\},
\nonumber\\
&& \Delta f_1(x_1,x_2)= \delta_{x_2,x^0_2} \bar f_1(x_1) ,
\nonumber\\
&& \Delta f_2(x_1,x_2)= f_2(x_1,x_2) \vert_{x_2=x^0_2-L/2-1} , 
\end{eqnarray}
where
$\bar f_1(x_1)= \sum_{y_2=-L/2}^{L/2-1}  f_1(x_1,y_2)$.

\vspace{1em}
IV.b(1-form):
\begin{eqnarray}
&& g_{12}(x_1,x_2)=- \sum_{y_2=x^0_2-L/2}^{x_2}
\left\{ f_1(x_1,y_2) - \delta_{y_2,x^0_2} \bar f_1(x_1) \right\}
\nonumber\\
&& \qquad \qquad \qquad
+  \sum_{y_1=x^0_1-L/2}^{x_1}
 f_2(y_1,x_2) \vert_{x_2=x^0_2-L/2-1}. \nonumber\\
 \end{eqnarray}


\begin{acknowledgments}
The authors would like to thank T.-W.~Chiu for the correspondence 
about the rational approximation of the overlap Dirac operator. 
The authors are also grateful to H.~Suzuki  for valuable discussions.
Y.K. is supported in part by Grant-in-Aid  for Scientific Research No. 14046207.
D.K. is supported in part by the Japan Society for Promotion of Science under 
the Predoctoral Research Program No. 15-887. 
\end{acknowledgments}


\begin{thebibliography}{99}
\bibitem{Ginsparg:1981bj}
P.~H.~Ginsparg and K.~G.~Wilson,
Phys.\ Rev.\ D {\bf 25}, 2649 (1982).


\bibitem{Neuberger:1997fp}
H.~Neuberger,
Phys.\ Lett.\ B {\bf 417}, 141 (1998)
[arXiv:hep-lat/9707022].

\bibitem{Hasenfratz:1998ri}
P.~Hasenfratz, V.~Laliena and F.~Niedermayer,
Phys.\ Lett.\ B {\bf 427}, 125 (1998)
[arXiv:hep-lat/9801021].

\bibitem{Neuberger:1998wv}
H.~Neuberger,
Phys.\ Lett.\ B {\bf 427}, 353 (1998)
[arXiv:hep-lat/9801031].

\bibitem{Hasenfratz:1998jp}
P.~Hasenfratz,
Nucl.\ Phys.\ B {\bf 525}, 401 (1998)
[arXiv:hep-lat/9802007].


\bibitem{Hernandez:1998et}
P.~Hernandez, K.~Jansen and M.~L\"uscher,
Nucl.\ Phys.\ B {\bf 552}, 363 (1999)
[arXiv:hep-lat/9808010].

%

\bibitem{Luscher:1998pq}
M.~L\"uscher,
Phys.\ Lett.\ B {\bf 428}, 342 (1998)
[arXiv:hep-lat/9802011].

\bibitem{Luscher:1998kn}
M.~L\"uscher,
Nucl.\ Phys.\ B {\bf 538}, 515 (1999)
[arXiv:hep-lat/9808021].

\bibitem{Luscher:1998du}
M.~L\"uscher,
Nucl.\ Phys.\ B {\bf 549}, 295 (1999)
[arXiv:hep-lat/9811032].

\bibitem{Luscher:1999un}
M.~L\"uscher,
Nucl.\ Phys.\ B {\bf 568}, 162 (2000)
[arXiv:hep-lat/9904009].

\bibitem{Luscher:1999mt}
M.~L\"uscher,
Nucl.\ Phys.\ Proc.\ Suppl.\  {\bf 83}, 34 (2000)
[arXiv:hep-lat/9909150].

\bibitem{Luscher:2000hn}
M.~L\"uscher,
arXiv:hep-th/0102028.


\bibitem{footnote:overlap}

Overlap formalism proposed by Narayanan and 
Neuberger\cite{
Narayanan:wx,Narayanan:sk,Narayanan:ss,Narayanan:1994gw,Narayanan:1993gq,
Neuberger:1999ry,Narayanan:1996cu,Huet:1996pw,
Narayanan:1997by,Kikukawa:1997qh} 
gives a well-defined partition function of Weyl fermions on the lattice,
which nicely reproduces the fermion zero mode and the fermion-number
violating observables
('t Hooft verteces)\cite{Narayanan:1996kz,Kikukawa:1997md,Kikukawa:1997dv}. 
Through the recent re-discovery of the Ginsparg-Wilson relation,
the meaning of the overlap formula, especially
the locality properties, become clear from the point of view 
of the path-integral. 
The gauge-invariant construction by L\"uscher\cite{Luscher:1998du} 
based on the Ginsparg-Wilson relation 
provides a procedure to determine the phase of the overlap formula
 in a gauge-invariant manner for anomaly-free chiral gauge theories.
For Dirac fermions, it provides a
gauge-covariant and local lattice Dirac operator
satisfying the Ginsparg-Wilson relation\cite{Ginsparg:1981bj,
Neuberger:1997fp,Kikukawa:1997qh,Neuberger:1998wv,
Hernandez:1998et}. The overlap formula was 
derived from the five-dimensional approach of 
domain wall fermion proposed by Kaplan\cite{Kaplan:1992bt}.
In its vector-like formalism\cite{Shamir:1993zy, Furman:ky,
Blum:1996jf, Blum:1997mz}, 
the local low energy effective action of the chiral mode 
precisely reproduces the overlap Dirac 
operator \cite{Vranas:1997da,Neuberger:1997bg, Kikukawa:1999sy}.


\bibitem{Narayanan:wx}
R.~Narayanan and H.~Neuberger,
Phys.\ Lett.\ B {\bf 302}, 62 (1993)
[arXiv:hep-lat/9212019].
%

\bibitem{Narayanan:sk}
R.~Narayanan and H.~Neuberger,
Nucl.\ Phys.\ B {\bf 412}, 574 (1994)
[arXiv:hep-lat/9307006].

\bibitem{Narayanan:ss}
R.~Narayanan and H.~Neuberger,
Phys.\ Rev.\ Lett.\  {\bf 71}, 3251 (1993)
[arXiv:hep-lat/9308011].

\bibitem{Narayanan:1994gw}
R.~Narayanan and H.~Neuberger,
Nucl.\ Phys.\ B {\bf 443}, 305 (1995)
[arXiv:hep-th/9411108].

\bibitem{Narayanan:1993gq}
R.~Narayanan,
Nucl.\ Phys.\ Proc.\ Suppl.\  {\bf 34}, 95 (1994)
[arXiv:hep-lat/9311014].

\bibitem{Neuberger:1999ry}
H.~Neuberger,
Nucl.\ Phys.\ Proc.\ Suppl.\  {\bf 83}, 67 (2000)
[arXiv:hep-lat/9909042].

\bibitem{Narayanan:1996cu}
R.~Narayanan and H.~Neuberger,
Nucl.\ Phys.\ B {\bf 477}, 521 (1996)
[arXiv:hep-th/9603204].

\bibitem{Huet:1996pw}
P.~Y.~Huet, R.~Narayanan and H.~Neuberger,
Phys.\ Lett.\ B {\bf 380}, 291 (1996)
[arXiv:hep-th/9602176].

\bibitem{Narayanan:1997by}
R.~Narayanan and J.~Nishimura,
Nucl.\ Phys.\ B {\bf 508}, 371 (1997)
[arXiv:hep-th/9703109].


\bibitem{Kikukawa:1997qh}
Y.~Kikukawa and H.~Neuberger,
Nucl.\ Phys.\ B {\bf 513}, 735 (1998)
[arXiv:hep-lat/9707016].

\bibitem{Narayanan:1996kz}
R.~Narayanan and H.~Neuberger,
Phys.\ Lett.\ B {\bf 393}, 360 (1997)
[Phys.\ Lett.\ B {\bf 402}, 320 (1997)]
[arXiv:hep-lat/9609031].

\bibitem{Kikukawa:1997md}
Y.~Kikukawa, R.~Narayanan and H.~Neuberger,
Phys.\ Lett.\ B {\bf 399}, 105 (1997)
[arXiv:hep-th/9701007].

\bibitem{Kikukawa:1997dv}
Y.~Kikukawa, R.~Narayanan and H.~Neuberger,
Phys.\ Rev.\ D {\bf 57}, 1233 (1998)
[arXiv:hep-lat/9705006].

\bibitem{Kaplan:1992bt}
D.~B.~Kaplan,
Phys.\ Lett.\ B {\bf 288}, 342 (1992)
[arXiv:hep-lat/9206013].

\bibitem{Shamir:1993zy}
Y.~Shamir,
Nucl.\ Phys.\ B {\bf 406}, 90 (1993)
[arXiv:hep-lat/9303005].

\bibitem{Furman:ky}
V.~Furman and Y.~Shamir,
Nucl.\ Phys.\ B {\bf 439}, 54 (1995)
[arXiv:hep-lat/9405004].

\bibitem{Blum:1996jf}
T.~Blum and A.~Soni,
Phys.\ Rev.\ D {\bf 56}, 174 (1997)
[arXiv:hep-lat/9611030].

\bibitem{Blum:1997mz}
T.~Blum and A.~Soni,
Phys.\ Rev.\ Lett.\  {\bf 79}, 3595 (1997)
[arXiv:hep-lat/9706023].

\bibitem{Vranas:1997da}
P.~M.~Vranas,
Phys.\ Rev.\ D {\bf 57}, 1415 (1998)
[arXiv:hep-lat/9705023].

\bibitem{Neuberger:1997bg}
H.~Neuberger,
Phys.\ Rev.\ D {\bf 57}, 5417 (1998)
[arXiv:hep-lat/9710089].

\bibitem{Kikukawa:1999sy}
Y.~Kikukawa and T.~Noguchi,
arXiv:hep-lat/9902022.

\bibitem{Fujiwara:1999fi}
T.~Fujiwara, H.~Suzuki and K.~Wu,
Nucl.\ Phys.\ B {\bf 569}, 643 (2000)
[arXiv:hep-lat/9906015].

\bibitem{Fujiwara:1999fj}
T.~Fujiwara, H.~Suzuki and K.~Wu,
Phys.\ Lett.\ B {\bf 463}, 63 (1999)
[arXiv:hep-lat/9906016].


\bibitem{footnote:noncompact-u1}
See also \cite{Neuberger:2000wq}
for a gauge-invariant construction of abelian 
chiral gauge theories in non-compact formulation.

\bibitem{footnote:nonabelian-anomaly}
For nonabelian chiral gauge theories, the local
cohomology problem can be formulated with 
the topological field in 4+2 dimensional
space.\cite{Luscher:1999un,
Luscher:1999mt,
Luscher:2000hn} So far, the exact cancellation of 
gauge anomaly has been shown in all orders of the perturbation
expansion for generic nonabelian theories\cite{Suzuki:2000ii, Igarashi:2000zi,
Luscher:2000zd},
and nonperturbatively for $SU(2)\times U(1)_Y$ electroweak 
theory, both in the infinite lattice\cite{Kikukawa:2000kd}. 
In the five-dimensional approach using the 
domain wall fermion\cite{Kaplan:1992bt,
Shamir:1993zy, Furman:ky,
Blum:1996jf, Blum:1997mz, 
Neuberger:1997bg, Kikukawa:1999sy}, the local cohomology problem
can be formulated in 5+1 dimensional space\cite{Kikukawa:2001mw}. 

\bibitem{Neuberger:2000wq}
H.~Neuberger,
Phys.\ Rev.\ D {\bf 63}, 014503 (2001)
[arXiv:hep-lat/0002032].

\bibitem{Suzuki:2000ii}
H.~Suzuki,
Nucl.\ Phys.\ B {\bf 585}, 471 (2000)
[arXiv:hep-lat/0002009].

\bibitem{Igarashi:2000zi}
H.~Igarashi, K.~Okuyama and H.~Suzuki,
arXiv:hep-lat/0012018.

\bibitem{Luscher:2000zd}
M.~L\"uscher,
JHEP {\bf 0006}, 028 (2000)
[arXiv:hep-lat/0006014].

\bibitem{Kikukawa:2000kd}
Y.~Kikukawa and Y.~Nakayama,
Nucl.\ Phys.\ B {\bf 597}, 519 (2001)
[arXiv:hep-lat/0005015].

\bibitem{Kikukawa:2001mw}
Y.~Kikukawa,
Phys.\ Rev.\ D {\bf 65}, 074504 (2002)
[arXiv:hep-lat/0105032].

\bibitem{Aoyama:1999hg}
T.~Aoyama and Y.~Kikukawa,
arXiv:hep-lat/9905003.

\bibitem{Kikukawa:1998pd}
Y.~Kikukawa and A.~Yamada,
Phys.\ Lett.\ B {\bf 448}, 265 (1999)
[arXiv:hep-lat/9806013].

\bibitem{Adams:1998eg}
D.~H.~Adams,
Annals Phys.\  {\bf 296}, 131 (2002)
[arXiv:hep-lat/9812003].

\bibitem{Fujikawa:1998if}
K.~Fujikawa,
Nucl.\ Phys.\ B {\bf 546}, 480 (1999)
[arXiv:hep-th/9811235].

\bibitem{Suzuki:1998yz}
H.~Suzuki,
Prog.\ Theor.\ Phys.\  {\bf 102}, 141 (1999)
[arXiv:hep-th/9812019].

\bibitem{Chiu:1998xf}
T.~W.~Chiu,
Phys.\ Lett.\ B {\bf 445}, 371 (1999)
[arXiv:hep-lat/9809013].

\bibitem{footnote:cohomology-at-finite-volume}
The cohomologically trivial part is, so far,
constructed in two steps:
the local cohomology problem is first solved in the
infinite lattice and then the corrections required
in a finite lattice are constructed and 
added\cite{Luscher:1999mt,Igarashi:2002zz}. 
Since the lattice Dirac operator satisfying the 
Ginsparg-Wilson relation should have the exponentially
decaying tail\cite{Horvath:1998cm,Horvath:1999bk}, 
the local fields in consideration should have the infinite 
number of components. Moreover, the vector potentials used in 
this analysis are not bounded.


\bibitem{Igarashi:2002zz}
H.~Igarashi, K.~Okuyama and H.~Suzuki,
arXiv:hep-lat/0206003.

\bibitem{Horvath:1998cm}
I.~Horvath,
Phys.\ Rev.\ Lett.\  {\bf 81}, 4063 (1998)
[arXiv:hep-lat/9808002].

\bibitem{Horvath:1999bk}
I.~Horvath,
Phys.\ Rev.\ D {\bf 60}, 034510 (1999)
[arXiv:hep-lat/9901014].


\bibitem{Kadoh:2003ii}
D.~Kadoh, Y.~Kikukawa and Y.~Nakayama,
``Solving the local cohomology problem in U(1) chiral gauge theories within a finite lattice,''
arXiv:hep-lat/0309022.


\bibitem{footnote:bound-neuberger}
It has been shown by Neuberger \cite{Neuberger:1999pz} that
the constant $1/30$ in the above bounds can be improved to
$1/6(2+\sqrt{2})$.


\bibitem{Neuberger:1999pz}
H.~Neuberger,
Phys.\ Rev.\ D {\bf 61}, 085015 (2000)
[arXiv:hep-lat/9911004].

\bibitem{Kikukawa:1998py}
Y.~Kikukawa and A.~Yamada,
Nucl.\ Phys.\ B {\bf 547}, 413 (1999)
[arXiv:hep-lat/9808026].

\bibitem{Neuberger:1998my}
H.~Neuberger,
Phys.\ Rev.\ Lett.\  {\bf 81}, 4060 (1998)
[arXiv:hep-lat/9806025].

\bibitem{Edwards:1998yw}
R.~G.~Edwards, U.~M.~Heller and R.~Narayanan,
Nucl.\ Phys.\ B {\bf 540}, 457 (1999)
[arXiv:hep-lat/9807017].

\bibitem{Chiu:2002eh}
T.~W.~Chiu, T.~H.~Hsieh, C.~H.~Huang and T.~R.~Huang,
Phys.\ Rev.\ D {\bf 66}, 114502 (2002)
[arXiv:hep-lat/0206007].

\bibitem{vandenEshof:2002ms}
J.~van den Eshof, A.~Frommer, T.~Lippert, K.~Schilling and H.~A.~van 
der Vorst,
Comput.\ Phys.\ Commun.\  {\bf 146}, 203 (2002)
[arXiv:hep-lat/0202025].


\bibitem{footnote:lemma-for-superlocal-field}
For the ultra-local tensor fields on a finite lattice,
the Poincar\'e lemma has been formulated by Fujiwara et al in 
\cite{Fujiwara:2000wn}.

\bibitem{Fujiwara:2000wn}
T.~Fujiwara, H.~Suzuki and K.~Wu,
Prog.\ Theor.\ Phys.\  {\bf 105}, 789 (2001)
[arXiv:hep-lat/0001029].

\bibitem{footnote:lemma-in-d}
An equivalent formulation of these lemma can be given
in terms of exterior difference operator $d$.

\bibitem{Fukaya:2003ph}
H.~Fukaya and T.~Onogi,
Phys.\ Rev.\ D {\bf 68}, 074503 (2003)
[arXiv:hep-lat/0305004].

\end{thebibliography}

\end{document}